\documentclass[preprint,3p,times]{elsarticle}
\usepackage{graphicx,xcolor,amsmath,amsthm,bm,xfrac,subfig}
\usepackage[utf8x]{inputenc}
\usepackage[T1]{fontenc}
\usepackage[english]{babel}
\usepackage{hyperref}
\hypersetup{colorlinks=true}

\biboptions{sort&compress}

\journal{Journal of Nuclear Materials}
\long\def\symbolfootnote[#1]#2{\begingroup%
\def\thefootnote{\fnsymbol{footnote}}\footnote[#1]{#2}\endgroup}
\long\def\symbolfootnotemark[#1]{\begingroup%
\def\thefootnote{\fnsymbol{footnote}}\footnotemark[#1]\endgroup}
\long\def\symbolfootnotetext[#1]#2{\begingroup%
\def\thefootnote{\fnsymbol{footnote}}\footnotetext[#1]{#2}\endgroup}

\begin{document}

\begin{frontmatter}
\title{Structural instability of the ground state of the U$_2$Mo compound}

\author[cab1]{E. L. ~Losada\corref{cor1}}
\author[cab2]{J.E.~Garc\'es}
\address[cab1]{SIM$^3$, Centro At\'omico Bariloche, Comisi\'on Nacional de Energ\'ia At\'omica, Argentina}
\address[cab2]{Gerencia de Investigaci\'on y Aplicaciones Nucleares, Comisi\'on Nacional de Energ\'ia At\'omica, Argentina}

\begin{abstract}
This work reports the structural instability at \textit{T=0} $^\circ\! K$ of the U$_2$Mo compound in the structure $C11_b$ under the distortion related to the $C_{66}$ elastic constant. The electronic properties of U$_2$Mo such density of states (DOS), bands and Fermi surface (FS) are studied to understand the source of the instability. The $C11_b$ structure can be interpreted as formed by parallel linear chains along the z-directions each one composed by successive U-Mo-U blocks. The hybridization due to electronic interactions inside the U-Mo-U blocks is slightly modified under the $D_6$ distortion. The change in distance among chains modifies the U-U interaction and produces the split of f-states. The distorted structure is stabilized by the energy lowering of the hybridized states, mainly between d-Mo and f-U states, together with the f-band split. Consequently, an induced Peierls distortion is produced in U$_2$Mo due to the $D_6$ distortion. In addition, the results of this work remarks that the structure of the ground state of the U$_2$Mo compound is other than the assumed $C11_b$ structure.

\end{abstract}

\end{frontmatter}
\symbolfootnotetext[1]{Corresponding author email: losada@cab.cnea.gov.ar}

\section{INTRODUCTION}\label{sec.introd}
\label{Intro}
The bcc U-Mo solid solution has technological relevance because is a promising candidate for developing high-density uranium fuel, to be used in research reactors with low enrichment uranium (LEU, 235U < 20 at\%) \cite{Snelgrove1997119}. It could be also a potential fuel  for transuranic-burning advanced reactors \cite{Kim2013520}. 

The development of high-density uranium fuels is one of the major challenges found in Material and Test Reactors research. Of particular interest is the U-Mo solid solution in the $\gamma$-phase dispersed in an Al matrix or in a monolithic form.
Both of them generated high expectation several years ago due to its apparent acceptable irradiation behaviour, low to moderate fuel/matrix interaction, and stable fission gas bubble growth for moderate neutron fluxes. However, unexpected failures like pillowing and large porosities in LEU UMo dispersion plates in high neutron flux irradiation experiments have been reported. Since these failures were found in 2004  \cite{VandenBerghe2008340,VandenBergheotro,Gan2010234}, much effort has been devoted to the search for their root causes and, mainly, to find a solution. It is now generally accepted that the properties of the interaction layer formed between the (U, Mo) fuel and the Al matrix are important factors in the failures. Consequently, the research has been focused in the development of diffusion barriers in the interface by adding additives to the fuel and/or to the matrix, mainly Si additions to the Al matrix. However, from the experimental point of view, it appears nowadays that the addition of Si in itself may be insufficient to meet the requirements of relatively high power at high burnup  \cite{VanderBerghe2012}. In addition, theoretical results using BFS methods for alloys showed that the formation of efficient diffusion barriers is strongly dependent on the Mo concentration near the interface \cite{GarcesHindawi}. In fact, the analysis shows that Si diffuses deeply in the U-Mo bulk for low levels of Mo in areas close to the interface. As the minimum necessary amount of Mo near the interface can not be controlled in a solid solution, an effective diffusion barrier can only be achieved locally in areas with enough Mo content near the interface. This theoretical result suggests that the possibility of obtaining a reliable and licensable fuel based on (Al,Si)/(U,Mo) could be a endless line of research \cite{GarcesSMTR}. The theoretical modelling also provides alternatives to the Si addition such as the C coating to produce an absolute diffusion barrier \cite{GarcesIGOR}. 
In any case, it is important to remark that the still unsuccessful development of (U, Mo) fuels leads to the need for other alternatives. Therefore experimental and theoretical research is needed to develop a reliable fuel. Besides the modelling effort using the BFS method, there is no other theoretical works known by the authors of this work aimed to describe the atomic behaviour in the interface. A few theoretical efforts using \textit{ab initio} methods have been made \cite{Landa2011132,Alonso2007,Jaroszewicz2013119} to understand the main thermodynamic properties of the U-Mo alloy and the stability of the U$_2$Mo compound in the $C11_b$ (MoSi$_2$ prototype) structure, which is the only stable compound observed in the phase diagram. The importance of \textit{ab initio} works is the possibility to describe the main physical properties of a system with no experimental information. However, the ground state must be properly characterized, mainly, if the cluster expansion technique is applied to obtain the correct behaviour of the disordered U-Mo alloy \cite{Sanchez1984334,Sluiter,Connolly}.
The work of Alonso et al. \cite{Alonso2007} studied the formation energy of the U-Mo solid solution in the bcc phase using the \textit{ab initio} cluster expansion technique with a set of effective cluster interactions (ECI). The ECI were computed using the direct inversion method from the total energies of 16 bcc-based ordered structures calculated by the Full-Potential Linearised Augmented Plane Wave method (FPLAPW)\cite{Wien2k}. These authors calculated the formation energy of the disordered $\gamma$ U–Mo solid solutions as well as the U$_2$Mo ($C11_b$) compound. The calculated energies of formation predict only one stable compound (U$_2$Mo) and an asymmetry that stabilizes the disordered alloy on the uranium rich side at $0^\circ $K. In addition, temperatures of calculated phase equilibrium using CVM were excessively high. The calculated disordering temperature of the U$_2$Mo ($C11_b$) compound is $\sim 2000 ^\circ $C, higher than the experimental value of $\sim 600 ^\circ $C \cite{Massalski2009}. This failure indicates that more clusters should be included in the cluster expansion and a bigger maximum cluster should be selected in the CVM method to study phase equilibria in U-Mo system. 

A more precise calculation of the heat of formation, the bcc ground state and phase diagram is presented by Landa et al. \cite{Landa2011132}. These authors applied the Green's-function technique based on the improved screened Korringa–Kohn–Rostoker method combined with the coherent potential approximation (CPA) to treat compositional disorder. In addition, the authors used the FPLMTO methods to compute the heat of formation of compounds and a 16-atom supercell to treat the compositional disorder within this formalism. Calculated heats of formation are compared with CALPHAD assessments. The decomposition curves for $\gamma$-based U-Mo solid solutions are derived from Ising-type Monte Carlo simulations. These authors discussed how the heat of formation correlates with charge transfer between the alloy components and found that the stabilization of the $C11_b$-phase in the U-Mo system is explained in terms of the electronic DOS change due to ordering of the U$_2$Mo alloy, i.e. a significant drop of the DOS in the vicinity of the Fermi level in the case of the ordered $C11_b$ compound causes a decrease of the total energy. Due to significant size mismatch between U and Mo atoms, these authors emphasize the importance of including the compositional and elastic dependence of ECI to produce lower temperature of decomposition of the $\gamma$ U-Mo solid solutions.

Remarkably, at the present time, only a few experimental data have been published related mainly to thermal and mechanical properties of the U$_2$Mo compound in the $C11_b$ structure \cite{Kutty2012193}. In addition, there is one theoretical attempt to model some structural, elastic and thermal properties of this compound using using density-functional theory (DFT) and the quasi-harmonic Debye theory \cite{Jaroszewicz2013119}. 

It is important to note here that all previous works assumed the $C11_b$ structure as the ground state of the U$_2$Mo compound. The results of this work show that this compound is unstable under the $D_6$ deformation related to the $C_{66}$ elastic constant. Consequently, the use of an unstable compound in the ECI expansion to describe the ground state of the U-Mo system could be the source of non-physical conclusions. This result reveals the importance of an accurate characterization of the physical and thermal properties of this compound and, mainly, the ground state of a system previously to apply the cited theoretical methodologies. 

The purpose of this work is to study the structural and electronic properties of U$_2$Mo to search for the origin of the instability under the $D_6$ deformation. The paper is organized as follows. Section \ref{sec.theory} gives the methodology used to compute the elastic constant. Section \ref{sec.results} presents the analysis of the structural instability observed in U$_2$Mo together with the electronic properties as DOS, bands and Fermi surface.

\section{THEORETICAL METHODOLOGY} \label{sec.theory}
\subsection{Structure optimization and calculation parameters}
The assumed crystal structure for the ground state of the U$_2$Mo compound is the body-centred tetragonal structure with the space group $I_4/mmm \; (\# 139)$, MoSi$_2$ type  prototype with  Strukturbericht designation $C_{11}b$ \cite{AYQ:AYQA01948}. The molybdenum atoms are at the positions 
 {$2(a)\!\!:$} $(0,0,0)$ and $(\sfrac{1}{2},\sfrac{1}{2},\sfrac{1}{2})$, and the uranium atoms at the positions {$4(e)\!\!:$} $(0,0,z)$, $(0,0,-z)$, $(\sfrac{1}{2},\sfrac{1}{2},\sfrac{1}{2}+z)$ and $(\sfrac{1}{2},\sfrac{1}{2},\sfrac{1}{2}-z)$ ; where $z=0.328 \pm 0.002 $. The experimental lattice parameters are $a_0=b_0=3.427 \AA$, $c_0=9.834 \AA$ \cite{Dwight196081}. 

\begin{figure}[h]
\center\includegraphics[scale=0.4]{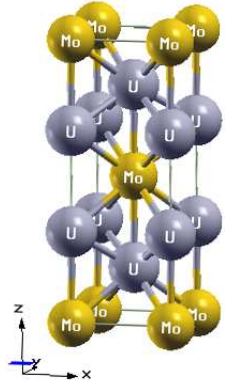}   
\caption{ The crystal structure of the assumed ground state for the U$_2$Mo compound \cite{Kokalj2003155,AYQ:AYQA01948}. \label{fig:U_2Mo}}
\end{figure}

The total energy of U$_2$Mo was computed using the full potential LAPW method based on density functional theory as it is implemented in the WIEN2k code \cite{Wien2k}. This code uses the full-potential LAPW+lo method that makes no shape approximation to the potential or density. The electronic exchange–correlation interactions were treated within the generalized gradient approximation of Perdew, Burke and Ernzerhof \cite{PhysRevLett.77.3865}, as no experimental or theoretical evidence of strong correlations was found on the system studied here. The radii of the atomic spheres ($R_{MT}$) selected for U and Mo were $R_{{MT}_{U}} = 2.4 \,a.u.$ and  $R_{{MT}_{Mo}} = 2.0 \,a.u.$.
In order to describe the electronic structure of all the atoms and its orbitals, in this calculation, the $APW+lo$ basis set was selected. This choice was made considering the fact this method allows to reach converged values at a lower Rkmax, and thus, the corresponding computational time is drastically reduced. Local orbital extensions were included to describe the semicore states by means of the method $(APW+lo)$ 
, as it is implemented in the WIEN2k program. The cutoff parameter that controls the convergence in the expansion of the solution to the Kohn–Sham equations is chosen to be $Rkmax = 8$, where kmax is the plane wave cutoff and RMT is the smallest of all the atomic sphere radii. The maximum $l$ values for partial waves used inside the atomic spheres and for the non-muffin-tin matrix elements were selected to be $lmax = 10$ and $lmax = 4$, respectively. The charge density cut-off Gmax was selected as $22 Ry^{1/2}$. A mesh of $10000$ k-points was taken in the whole Brillouin zone. k-space integration is done by the modified tetrahedron-method \cite{PhysRevB.49.16223}. The iteration process is repeated until the calculated total energy converges to less than $1x10^{-6} Ry/cell$, and the calculated total charge converges to less than $1x10-6 e/cell$. A complete determination of the tetragonal structure requires the determination of three parameters (a, b and the internal position parameter). In order to calculate the internal parameters of the crystal structure we employed the mini LAPW script implemented in the Wien2K package .

\begin{figure}[tp]
\center\includegraphics[]{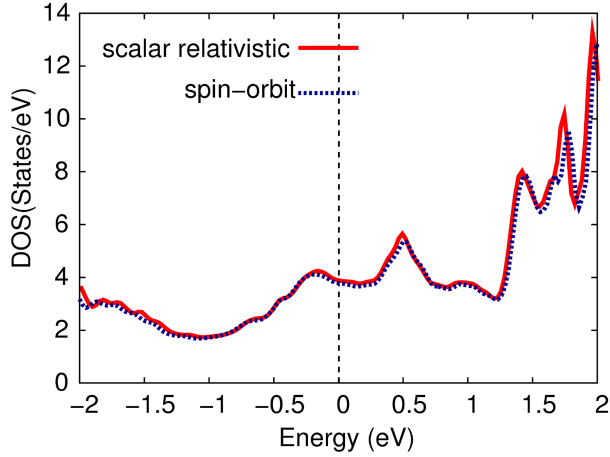}  
\caption{Comparison of the density of states around $E_F$ with and without spin orbit. k-points=$10000$. } \label{compDOS}
\end{figure}

The theoretical lattice parameters at $0 ^\circ$K obtained were $a=b=3.4437 \AA$, $c=9.6842 \AA$ and $z=0.3216$, in good agreement with experimental results. The calculations were performed without including spin-orbit interactions because no significant effects were observed near the Fermi level, as it is shown in Fig. \ref{compDOS} for the total DOS.

\subsection{Elastic constants $C_{44}$ and $C_{66}$ calculations in U$_2$Mo} \label{elastic}

The elastic constants are obtained by computing the second order derivatives of the total energy with respect to the distortion parameter. The elastic constants of $U_2Mo$ were studied in a previous work \cite{Jaroszewicz2013119}, using the density functional theory implemented in the WIEN2k code. These calculations were performed by using a package included in the code which calculates elastic tensors of tetragonal phases \cite{Reshak2012147}. The results are presented in Table \ref{tb:econstant}.

\begin{table}[h]
\caption{Calculated elastic constants (in GPa), bulk modulus B, Young's modulus E, shear modulus G and Poisson's ratio $\nu$. }\label{tb:econstant}
\vspace{-0.5cm}
\begin{center}
\begin{tabular*}{0.4\textwidth}{@{\hspace{6 mm}} c @{\hspace{6 mm}} c @{\hspace{1.4 cm}} l @{\hspace{6 mm}} c }
\hline
 $C_{11}$ & $254$ & B & 180 GPa\\
 $C_{12}$ & $161$ & E & 102 GPa\\
 $C_{13}$ & $125$  & G & 36 Gpa\\
 $C_{33}$ & $295$ & $\nu$ & 0.4\\
 $C_{44}$ & $38$  & \\
 $C_{66}$ & $\sim 20$ \\ 
 \hline
\end{tabular*}
\end{center}
\vspace{-2 mm}
\end{table}

\begin{figure}[h]
\center\includegraphics[scale=0.7]{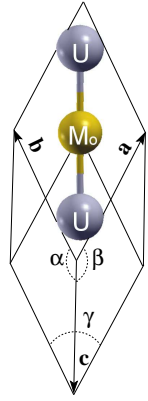}   
\caption{U$_2$Mo primitive unit cell. It is a triclinic crystal structure with vector modules: $\vert\mathbf{a}\vert =\vert\mathbf{b}\vert =\vert\mathbf{c}\vert =5.4199\AA$ and angles: $\alpha =\beta =142.953766 ^{\circ}$, $\gamma =53.394693^{\circ}$. \label{fig:prim}}
\end{figure}

The elastic constants are very sensitive to the data point selection due to numerical inaccuracies of the calculations around the minimum of the total energy. In order to have a better control of the outcome in the cases of constants with very small values, such as $C_{44}$ and $C_{66}$, the calculations of the total energies were repeated using the Wien2k code but not its elastic packages. This was achieved by considering the structure in the primitive triclinic unit cell with space group P1 (see Fig.\ref{fig:prim} ). After global relaxation, the primitive lattice vector are $a=b=c= 5.4199 \AA$ and angles $\alpha =\beta= 142.954 ^\circ $ and $\gamma= 53.395 ^\circ $.  The basis comprises only three atoms, one Molybdenum at position $(0 0 0)$, one Uranium at $(u\;u\;0)$ and the other Uranium at $(\bar{u}\;\bar{u}\;0)$; being $u=0.3216$. All the positions are referred to the primitive lattice vectors $\vec{a}$, $\vec{b}$ and $\vec{c}$ in crystal parameters. See Fig. \ref{fig:prim}. 

The elastic constants $C_{44}$ and $C_{66}$ are calculated through the volume conserving monoclinic shear distortions $D_4$ and $D_6$, respectively. Matrices for each distortion are given by Eq. \ref{eq:d4} and \ref{eq:$D_6$}, 

\begin{align}\label{eq:d4}
D_4=\dfrac{1}{(1-\delta ^2)^{\sfrac{1}{3}}}\left(
\begin{array}{ccc}
1&0&0\\
0&1&\delta\\
0&\delta &1\\
\end{array} 
\right) \,
\end{align}

\begin{align}\label{eq:$D_6$}
D_6=\dfrac{1}{(1-\delta ^2)^{\sfrac{1}{3}}}\left(
\begin{array}{ccc}
1&\delta&0\\
\delta&1&0\\
0&0&1\\
\end{array} 
\right) \,
\end{align}
where, $\delta$ is the dimensionless distortion parameter. 

The energy associated to this distortion can be written in the following way:
\begin{align}
E(V,\delta)=E(V_0,0)+V_0(2\tau _4\delta +2c_{44}\delta ^2)\,,\\
\nonumber\\
E(V,\delta)=E(V_0,0)+V_0(2\tau _6\delta +2c_{66}\delta ^2)\,,
\end{align}
where $V_0$ is the volume of the unstrained system, $E(V_0,\delta)$ is the total energy computed by \textit{ab initio} codes in this work, $\tau _4$ and $\tau _6$ are the elements of the stress tensor in Voigt notation ($ \tau _4= \tau _{yz}$, $ \tau _6= \tau _{xy}$). Therefore, the elastic constants are deduced from the following expressions:

\begin{align}
c_{44}=\dfrac{1}{4V_0}\left. \dfrac{\partial ^2 E}{\partial \delta ^2}\right\vert _V \,,\\
\nonumber\\
c_{66}=\dfrac{1}{4V_0}\left. \dfrac{\partial ^2 E}{\partial \delta ^2}\right\vert _V \,.
\end{align}

\section{RESULTS AND DISCUSSIONS} \label{sec.results}
\subsection{Structural instability in $U_2Mo$}

\begin{figure}[h]
\centering
\includegraphics[]{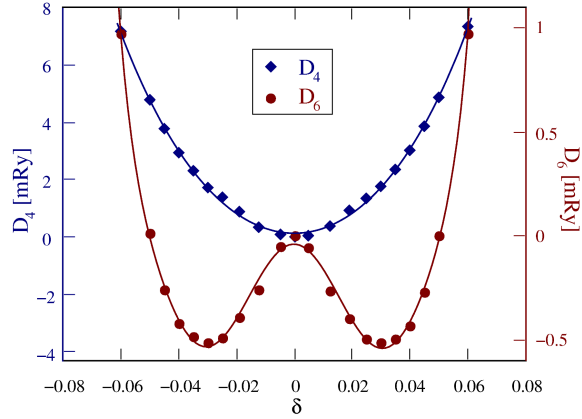}
\caption{Total energy for different values of the distortion parameter $\delta$ in U$_2$Mo after the relaxation of internal positions. Spin-orbit coupling was not included in the calculations. \label{fig:$D_6$YD4}}
\end{figure}

The low magnitude of the $C_{44}$ and $C_{66}$ constants led the question whether a soft phonon mode is related to those deformations or if a more precise calculation around the minimum of energy could reveal hidden structural anomalies due to strong electron-phonon coupling.\newline
Following the procedure outlined in section \ref{sec.theory}, a lattice instability was found for the $D_6$ deformation but none for the $D_4$ one. Fig. \ref{fig:$D_6$YD4} shows the total energy versus the distortion $\delta$ for the compound U$_2$Mo in $C11_b$ structure. The total energy reaches a minimum value for a displacement of $\delta = \pm 0.03$. The magnitude and shape of the total energy vs. distortion curve obtained under the application of the $D_6$ distortion in U$_2$Mo is very similar to the one calculated in the $\alpha$-U system, where a charge density wave is observed \cite{PhysRevLett.81.2978}. However, whereas the CDW in $\alpha$-U is spontaneous, the effect observed in U$_2$Mo is induced by deformation and could be a remnant of the effect observed in pure U. If so, U$_2$Mo will be another example of a three-dimensional system with this kind of phenomenology. In addition, the instability remarks that the ground state of U$_2$Mo compound is a crystalline structure different to the $C11_b$.

\subsection{$D_{6}$ distortion in U$_2$X (X= Tc, Rh, Ru, Cr, Pd, W, Al, Nb, Z) metastable compounds}

\begin{figure}[h]
\centering
\subfloat[X element in the same period (5).]{\label{fig:WHoriz}\includegraphics[width=0.4\textwidth]{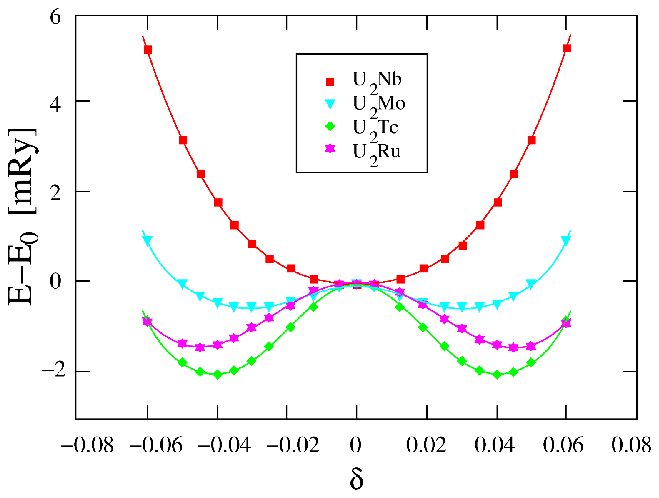}}
\hspace*{1cm}
\subfloat[X element in the same group (VI).]{\label{fig:WVert}\includegraphics[width=0.4\textwidth]{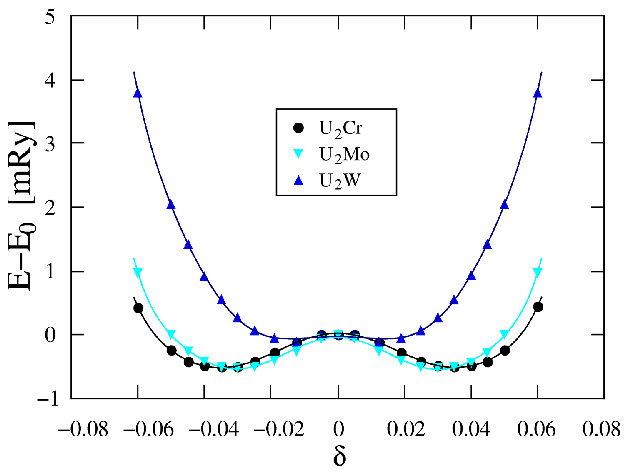}}
\caption{Comparison of the total energy variation under application of the $D_6$ distortion parameter $\delta$ for different $U_2X$ compounds.  $E_0$ corresponds to the value of energy at the undistorted lattice.}\label{fig:Wotros}
\end{figure}

In order to study the effect produced by the occupation of the electronic states on the structural instability related to the $D_6$ distortion, the behaviour of the original U$_2$Mo structure was analysed by replacing the Mo atoms with U atoms (U$_2$U) and also with other nearby elements of the periodic table (Tc, Rh, Ru, Cr, Pd, W, Al, Nb, Zr). The calculations will give information whether it is a specific effect related to only the U$_2$Mo compound or is a more general effect involving other elements with different occupation of the d-states. The $D_6$ distortion matrix was applied to the unit cell (see Fig. \ref{fig:prim}) after volume optimization and internal parameters relaxation of each structure in the space group $I4/mmm$. The total energy differences relative to the values obtained for the undistorted cells as a function of the distortion parameter $\delta$, are shown in Fig. \ref{fig:WHoriz} and \ref{fig:WVert}. The former compares elements situated in the same period (5) and the later one compares elements on the same group (VI).  

Whereas it is not observed any instability in the U$_2$U compound, the existence of such instability is observed in other U$_2$X compounds. The result emphasizes two facts: i) the U atoms involved in the CDW in $\alpha$-U behave different in the U$_2$X compounds and, ii) the interaction and number of d-electrons play a crucial role in the existence of the instability. 
Figs. \ref{fig:WHoriz} and \ref{fig:WVert} show that the structural instability is observed only if the transition elements have at least five d-electrons. The results reveal that the interaction between d- and f-electrons plays a fundamental role in the instability. Therefore, the effects of the hybridization should be observed in the electronic properties such as DOS, band and Fermi surface.

\subsection{Electronic properties analysis}
\subsubsection{DOS and projected-DOS}

\begin{figure}[h]
\centering
\subfloat[$\delta =0$]{\label{fig:Dos-0}\includegraphics[width=0.4\textwidth]{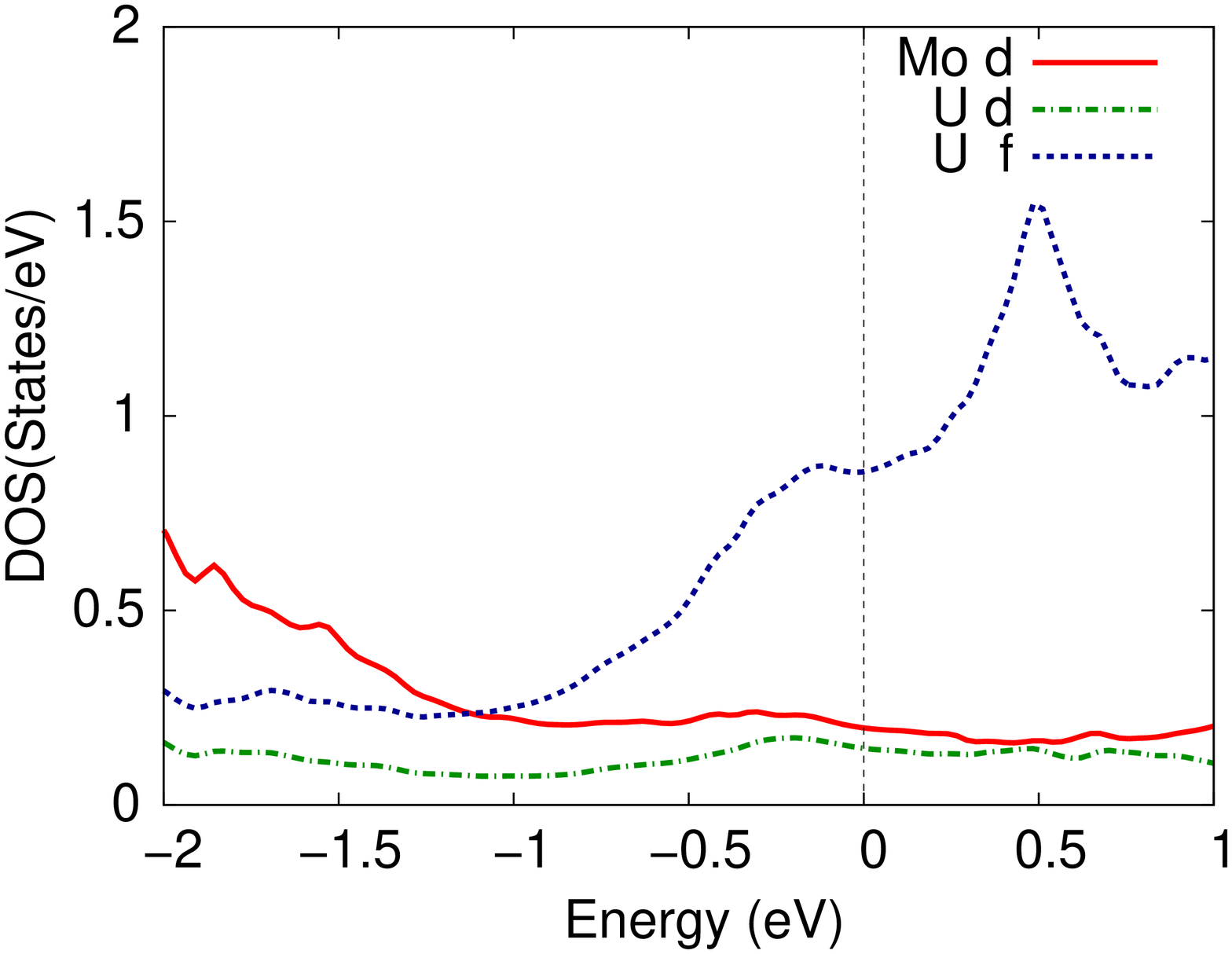}}
\hspace*{1cm}
\subfloat[$\delta =0.03$]{\label{fig:Dos-3}\includegraphics[width=0.4\textwidth]{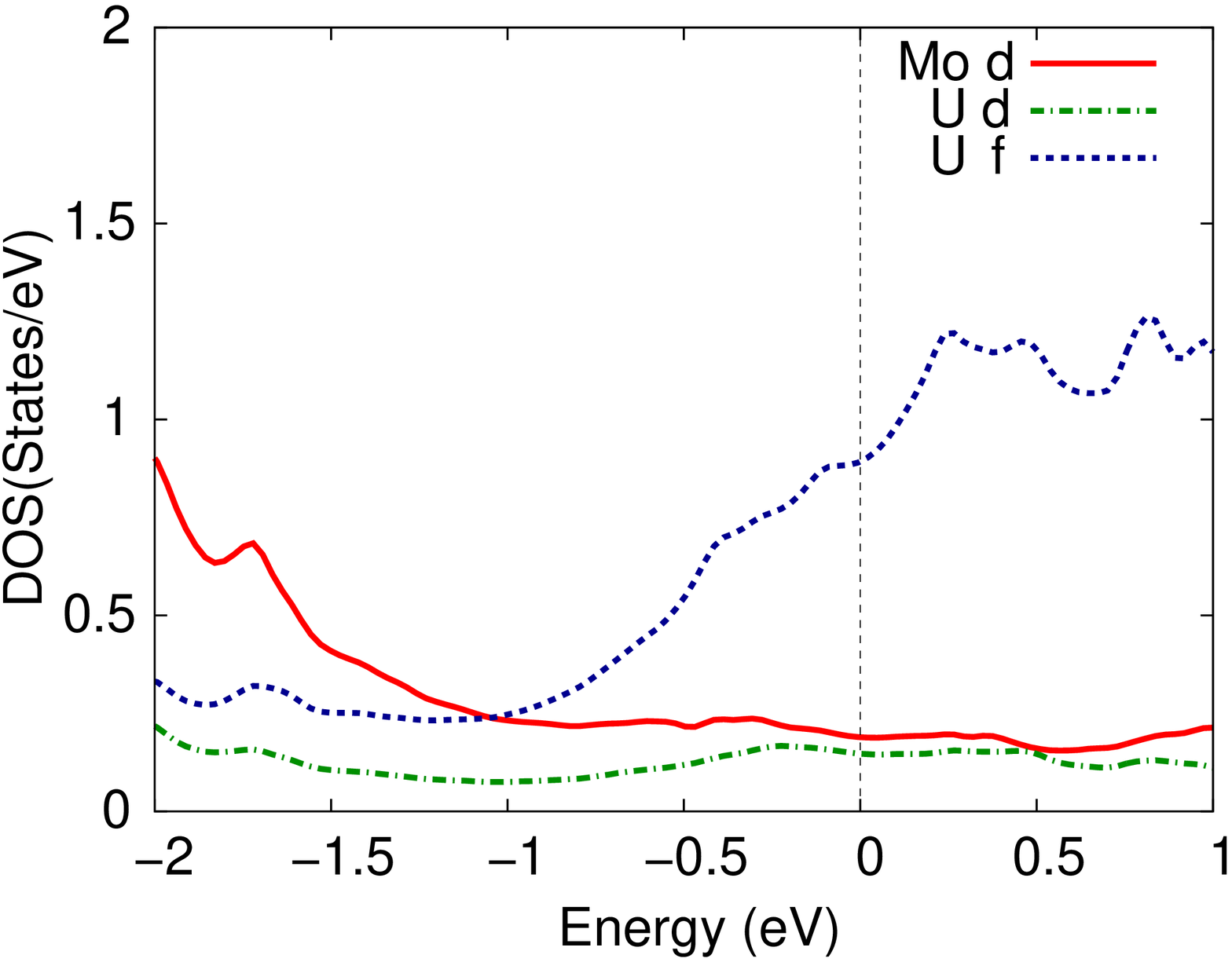}}\\ 
\caption{Electronic density of states and projections onto the LAPW spheres. k-points=$50000$}\label{fig:Dos}
\end{figure}

The first electronic properties computed were the DOS and partial DOS. These calculations were done relativistically, without including spin-orbit interaction, as spin-orbit does not seem to have a large effect, as it was shown in  Fig. \ref{compDOS}. A comparison of the partial DOS is given in Fig.\ref{fig:Dos-0} for the undistorted ($\delta$=0) and in Fig.\ref{fig:Dos-3} for the distorted ($\delta$=0.03) cases. A small change of the projected f-DOS is observed below and around the Fermi level. It is clear from the projected DOS that the f-states of U play a fundamental role near the Fermi level (FL), similarly to $\alpha$-Uranium.

\subsubsection{Electronic energy band structure}
\begin{figure}[h]
\centering
\includegraphics[scale=0.4]{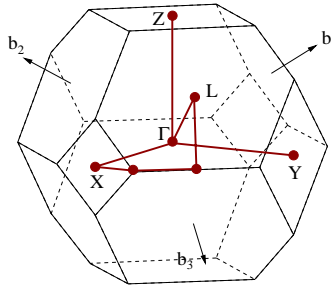}%
\caption{Brillouin zone of $ORCF_2$ (Face-centred orthorhombic $1/a^2 < 1/b^2 + 1/c^2  $)  type lattice \cite{Setyawan2010299} with the selected path for bands calculations.}\label{fig:path}
\end{figure}

\begin{figure}[h]
\centering
\subfloat[$\delta =0$]{\label{fig:bands0}\includegraphics[width=\textwidth]{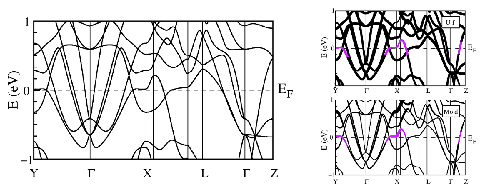}} \\
\subfloat[$\delta =-0.03$]{\label{fig:bands3}\includegraphics[width=\textwidth]{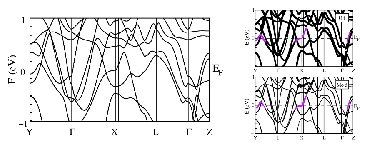}}
\caption{Band structure (left) and predominant character acting nearby the Fermi level (right) obtained following the path: $Y-\Gamma - X -L -\Gamma - Z $k-points=$50000$.}\label{fig:bands}
\end{figure}

The electronic band structure for U$_2$Mo was calculated through the path shown in Fig. \ref{fig:path}. The orbital character is illustrated in Figs. \ref{fig:bands0} and \ref{fig:bands3} for the undistorted and distorted structure, respectively. The f-states of U and the d-states of Mo are plotted in a 1 eV range around EF.

The bands plot for the undistorted structure shows the Fermi level (FL) intersected mainly by the f-states of U. However, there are zones near the FL showing d-states of Mo atoms hybridized with the f-states. They are located near the X, Y and Z points of the Brillouin zone. They are coloured in purple in Fig. \ref{fig:bands}. 
For the distorted case, it is observed in Fig. \ref{fig:bands3} an important rearrangement of the bands. The lowered symmetry of the distorted structure breaks the degeneracy of the f-bands of U. There is an important split of the bands just below the FL near the $\Gamma$ point. The main splits are observed in the directions $\Gamma$-X, $\Gamma$-Y, $\Gamma$-Z and $\Gamma$-L. Those changes modify completely the number of states crossing the FL modifying the structure of the Fermi surface. It is still observed the hybridization of d-Mo and f-U states, but 
those states move below the Fermi level near the Y, X and Z points. Consequently, there are occupied states in the distorted structure with a lower energy than the same states of the symmetric structure. Therefore, the low symmetry structure is stabilized by the relocation below the FL of the hybridized states between Mo and U atoms together with the f-band split of U, as shown in Fig. \ref{fig:bands3}. At variance with $\alpha$–U, the d-sates of Molybdenum play an important role in stabilizing the distorted structure. 
Changes like those observed in the bands of the distorted U$_2$Mo, stabilizing structures with lower symmetry, are usually referred to as a Peierls distortion. However, whereas the Peierls distortion in $\alpha$-U is spontaneous, in U$_2$Mo is induced by the deformation $D_6$.

The $C11_b$ structure  can be interpreted as formed by parallel linear chains along the z-directions each one composed by U-Mo-U blocks. The changes due to the distortion $D_6$ modify the distance between the Mo and U atoms. The hybridization mainly due to interaction inside the U-Mo-U blocks is slightly modified. Those states just move below the Fermi level contributing to stabilize the distorted structure. The changes in distance among chains modify the U-U interaction and reproduce the same situation observed in $\alpha$-U, i.e. the split of f-state. The induced Peierls distortion in U$_2$Mo arises from this mechanism.

\begin{figure}[h!]
\centering
\subfloat[Band \# 30]{\label{fig:fs030}\includegraphics[width=0.17\textwidth]{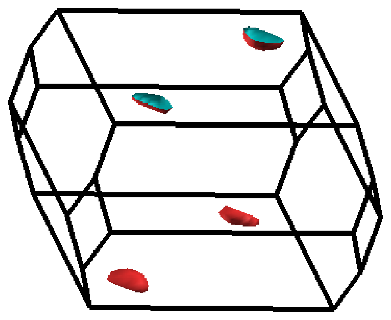}} 
\subfloat[Band \# 31]{\label{fig:fs031}\includegraphics[width=0.17\textwidth]{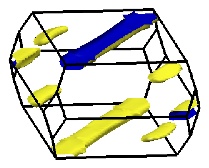}} 
\subfloat[Band \# 32]{\label{fig:fs032}\includegraphics[width=0.17\textwidth]{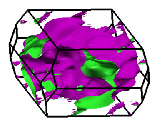}}
\subfloat[Band \# 33]{\label{fig:fs033}\includegraphics[width=0.17\textwidth]{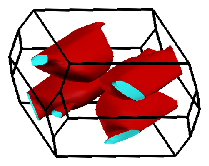}}\\
\caption{Fermi surfaces corresponding to the different intersecting bands with $\delta =0$. k-points=$50000$.}\label{fig:fs0}
\centering
\subfloat[Band \# 31]{\label{fig:fs331}\includegraphics[width=0.17\textwidth]{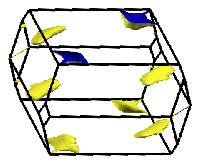}} 
\subfloat[Band \# 32]{\label{fig:fs332}\includegraphics[width=0.17\textwidth]{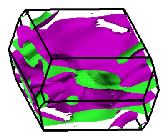}}
\subfloat[Band \# 33]{\label{fig:fs333}\includegraphics[width=0.17\textwidth]{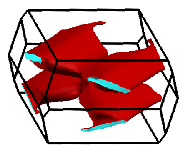}}\\
\caption{Fermi surfaces corresponding to the different intersecting bands with $\delta =-0.03$. k-points=$50000$.}\label{fig:fs-3}
\end{figure}

\vspace{-0.5cm}
\subsubsection{Fermi surface}
The Fermi surfaces for the undistorted and distorted systems are shown in Figs. \ref{fig:fs0} and \ref{fig:fs-3} respectively. There is a clear topological change in the Fermi surface close to the X point, and mainly, in the $\Gamma$-X, $\Gamma$-Y and $\Gamma$-Z directions. 
The band \# 30 has disappeared in the distorted structure. In addition, the band \# 31 becomes disconnected along the $\Gamma$-X direction. The states along the other two directions produce a connected Fermi surface, observed clearly in band \# 33.  The mechanism is similar to the Peierls distortion that occurs in systems with narrow bands intersecting the Fermi level, as was explained by Soderlind for $\alpha$-U \cite{Soderlind2002}. Both conditions, necessary for the existence of a charge density wave, are met in the case of the structure of the compound $C11_b$. 

\section{Conclusions}

Besides the technological importance of the U-Mo solid solution in nuclear energy research, there are few theoretical attempts to model this system. All of previous works assumed the $C11_b$ structure as the ground state of the U$_2$Mo compound. This work reports the existence of a structural instability of this compound under the deformation $D_6$, related to the $C_{66}$ elastic constant. This result will be helpful in the modelling of the U-Mo system with theoretical methods using ground state information.
The structural and electronic properties of U$_2$Mo are studied in this work to find the source of the instability under the $D_6$ deformation. It is found that the split of the f-states of U together with the energy lowering of the hybridized states, mainly between d-Mo and f-U states, are responsible of the stabilization of the distorted structure. The Fermi surface is analysed. It is found a clear topological change in the Fermi surface close to the X point, and mainly, in the $\Gamma$-X, $\Gamma$-Y and $\Gamma$-Z directions. 
Changes like those observed in the bands of the distorted U$_2$Mo, i.e. the split of f-bands and the lowering of f-states, stabilizing structures with lower symmetry, are usually referred to as a Peierls distortion. However, whereas the Peierls distortion in $\alpha$-U is spontaneous, in U$_2$Mo is induced by the deformation $D_6$.
It remains to study whether a connection between the $D_6$ deformation and a CDW can be established and to find the structure of the ground state of the U$_2$Mo compound.

\bibliographystyle{apsrev4-1}
\bibliography{bibliografia.bib}

\end{document}